\documentclass[twocolumn,preprintnumbers,amsmath,amssymb]{revtex4}

\usepackage{graphicx}
\usepackage{amssymb}
\usepackage{psfrag}
\def\e{\eta}

\def\l{\lambda}
\def\t{\theta}
\def\c{\chi}

\newcommand\beq{\begin{equation}}
\newcommand\eeq{\end{equation}}
\newcommand\bea{\begin{eqnarray}}
\newcommand\eea {\end{eqnarray}}

\begin{document}

\title{Rotation of galaxies as a signature of cosmic strings in weak lensing surveys}

\author{Daniel B.~Thomas, Carlo R.~Contaldi, Jo$\tilde{\textrm{a}}$o Magueijo}
\affiliation{Theoretical Physics, Blackett Laboratory, Imperial College, London}
\date{\today}

\begin{abstract}
  Vector perturbations sourced by topological defects can generate
  rotations in the lensing of background galaxies. This is a potential
  smoking gun for the existence of defects since rotation generates
  a curl-like component in the weak lensing signal which is not
  generated by standard density perturbations at linear order. This rotation
  signal is calculated as generated by cosmic strings. Future
  large scale weak lensing surveys should be able to detect this
  signal even for string tensions an order of magnitude lower than
  current constraints.
\end{abstract}

\maketitle


{\it Introduction.} Weak lensing of background galaxies 
has earned a place in the growing
observational toolkit of the era of precision cosmology. The shearing
of galaxies out to redshifts of a few has become a routine 
measurement~\cite{refregier03}. 
These measurements hold much promise in the quest to
constrain cosmological parameters with a particular focus on the
equation of state of dark energy \cite{lsstdark,panstarrs}. The conventional
picture first proposed in \cite{kaiser92} is one where correlations in
the weak lensing of photon bundles can be related statistically to the
power spectrum of the evolving density field along the line of
sight. The correlations can be observed by measuring the shearing of
background objects such as galaxies or higher redshift objects such as
Ly-$\alpha$ emitters or CMB anisotropies.

The statistical effect of the weak lensing is well understood. Scalar
perturbations such as density fluctuations  generate three
independent components of the matrix relating the original
source to the distorted image. The first is a trace $\kappa$ which gives
the amplification, or convergence of the image and the second are two
shear components, $\gamma_1$ and $\gamma_2$, which describe a
symmetric, traceless, and divergenceless contribution to the
distortion matrix. 
A fourth independent component $\rho$, describing rotations, can be added
as an anti-symmetric contribution. However, $\rho$ cannot be generated
by linear perturbations transforming as scalars under 3d rotations. In
fact any rotational, or curl-like, component in the surveys has been
used as measure of systematic contamination of the data \cite{cfhtls}. 

A number of authors \cite{baconwl,coorayhu,sarkarwl} have extended the formalism to
account for the fact that scalar density perturbations can source
$\rho$ via generation of vector (bulk flows) and tensor (gravity waves)
at second order. In both cases however, the signal is expected to be
very small and it will be a significant challenge to measure even with
future surveys. Another source is intrinsic correlations in 
galaxies~\cite{penintrinsic,leeintrinsic}.

This {\sl letter} suggests an alternative source of curl-like
distortions at first order in the perturbation amplitude. The source
of the signal are vector metric perturbations induced by cosmic
strings along the line of sight. Cosmic strings were first predicted
in the context of symmetry breaking phase transitions in the early
universe \cite{kibble}. They arise as topological defects along lines
where a complex field has remained trapped in a false vacuum after a
symmetry breaking phase transition where the field rolls down to a
global vacuum selecting a random phase. 

For many years cosmic strings provided an ``active'' alternative to the
``passive'' structure formation scenarios based on inflationary generated
passive perturbations (the terminology originates in~\cite{Magueijo96}). 
The passive picture became accepted as the
predominant mechanism when the first acoustics peaks, a clear prediction
of the coherent passive scenario, were detected in the CMB
\cite{sask,boom}. However a sub-dominant contribution from cosmic
strings has not been ruled out \cite{bevis1,battyecmb}. In recent years renewed
interest in cosmic strings has also been driven by the possibility
that many string theory models predict the generation of
macroscopic strings at the end of inflation \cite{cosmicsstring}.

Cosmic strings carry energy and momentum and source perturbations
to the metric. The metric perturbations in turn lead to lensing
of photon trajectories close to the strings. This is the source of the
well known Kaiser--Stebbins effect \cite{ks,ks2} where a moving string
causes a line--like discontinuity in the CMB
temperature. The signal induced in cosmic shear surveys
by the scalar source of a network of strings is smaller than that
due to dark matter density perturbations along the line of sight given
current constraints on string tensions $G\mu < 0.7 \times 10^{-6}$ (for 
the Abelian model~\cite{bevis1}). 
In contrast, the signal due to vector and tensor
perturbations sourced by strings generates rotations which
have no counterpart, at the same order of magnitude, from density
perturbations. Thus any observations of curl--like lensing signal would
provide a candidate detection of cosmic strings.   

This {\sl letter} focuses on the vector mode induced signal which is
expected to be an order of magnitude greater than the tensor induced
one \cite{contaldi99}. The {\sl letter} is organized as follows; the
lensing distortion generated by a vector source is first calculated
and then applied to the case of a single, moving, straight
string. Finally the statistical signal due to a network
of strings is computed and compared with the expected variance of
future weak lensing surveys. Throughout units where $c=1$ are used unless otherwise
stated. Greek indices run over all spacetime dimensions with latin
indices running only over the spatial dimensions. Overdots denote
differentiation with respect to conformal time $\eta$ and a 
$(-\, +\, +\, +)$ signature is adopted for the metric.

{\it Vector sourced distortions.} Generalised, vector--type
perturbations to the flat Friedmann-Robertson-Walker metric are given
by the contributions $g_{0i}=-a^2V_i$ and
$g_{ij}=a^2(F_{i,j}+F_{j,i})$, with both $F_i$ and $V_i$ are
divergenceless vectors and $a(\eta)$ is the scale factor. Two of the
four independent modes specified by the two vectors can be fixed by a
choice of gauge and $F_i=0$ is adopted for this calculation. The
geodesic equation can then be used to derive the effect of the perturbed metric on the
trajectory of photons \cite{dodelson}.
The coordinates can be aligned such that $x^i=(x,y,z)\equiv\c(\t_1,\t_2,1)$ where $\chi$
is the comoving radial distance with $d\chi/d\eta =1$ and $\vec
\theta$ is the vector spanning the plane orthogonal to the line of
sight. Using the relation $d\eta/d\l=p/a$, where $p$ is the modulus of
the photon 3-momentum, a second order differential equation
for the transverse projection of the trajectory is obtained
\begin{equation}
\frac{d^2(\c\t_i)}{d\c^2}=\frac{\dot V _i}{a^2}+\frac{V_{z,i}}{a^2}-\frac{V_{i,z}}{a^2}\text{,}
\end{equation}
where $i=1$ and $2$ only. In the small angle approximation the
transverse deflection can be derived as a $2 \times 2$ jacobian matrix
relating the observed source position $\theta_i$ to its true position
on the transverse source plane $\theta^{\prime}_j$ as $\partial \t^\prime_i/\partial \t_j=\delta_{ij}+\psi_{ij}$ such that 
\begin{equation}
\psi_{ij}\!=\int^{\c_{\infty}}_0d\c \ g(\c)\left(\frac{\dot V _{i,j}
    +V_{z,ij}-V_{i,zj}}{a^2}\right)
\end{equation}
with $ g(\c) \equiv
\c\int^{\c_{\infty}}_{\c}d{\tilde \c}\left(1-\c/{\tilde \c}\right)W({\tilde \c})$
a weighted integral of the normalised source distribution function $W(\chi)$
along the line of sight.

In the case examined here, the metric perturbations $V_i$ are sourced
by vector perturbations in the cosmic string stress--energy
tensor. These are described in terms of a divergenceless vector
contribution to the string momentum $\omega_i$ and a divergenceless
and traceless contribution to the anisotropic stress $\Pi_i$. The
sources are coupled to the metric perturbation via the Einstein equations
\begin{eqnarray}
  V_i&=&\frac{16\pi G a^2}{k^2}\omega_i \nonumber \\
  \dot{V} _i&=&-\frac{8\pi G a^2 \Pi _i}{k}-\frac{2\dot a}{a}\left( \frac{16\pi G a^2}{k^2}\right)\omega_i \text{,}
\end{eqnarray}
where the perturbations have been implicitly expanded in 3d plane waves
$\exp(-i{\vec k}\cdot{\vec x})$. $V_i$ and $\dot
V_i$ can then be eliminated to
obtain the distortion tensor $\psi_{ij}$ in terms of the vector
sources ${ \omega}_i$ and ${ \Pi}_i$
\begin{eqnarray}\label{psi}
\psi_{ij}&=&\frac{2G}{\pi^2}\int^{\c_{\infty}}_0d\c
g(\c)\!\int^{\infty}_{-\infty}d^3ke^{i\vec{k}\cdot\vec{x}} \times\nonumber\\
&&\hat{k}_j\!\left(\! \hat{k}_i \omega_z-\hat{k}_z \omega_i-2\frac{\dot{a}}{a} \frac{\omega_i}{k} -\frac{1}{2}\Pi_i\!\right)\text{,}
\end{eqnarray}
where $\hat k _i\equiv k_i/|\vec k|$. The convergence, shear, and rotation modes can then be inferred from
the distortion tensor using
\begin{equation}
  - {\bf \psi}_{ij} \equiv \left( \begin{array}{cc} \kappa+\gamma_1 & \gamma_2 +
    \rho \\ \gamma_2 - \rho & \kappa - \gamma_1\end{array} \right)\text{.}
\end{equation}

\begin{figure}[t]
\begin{center}
\psfrag{zaxis}{$y\equiv \chi \theta_2$}
\psfrag{xaxis}{$x\equiv \chi \theta_1$}
\psfrag{yaxis}{$z\equiv \chi$}
\psfrag{vel}{$\vec v$}
\psfrag{chi1}{$\chi=\chi_{cs}$}
\psfrag{chi2}{$\chi=\chi_g$}
\psfrag{pos1}{$\chi \vec \theta^{\, \prime}$}
\psfrag{pos2}{$\chi \vec \theta$}
\psfrag{string}{$\mbox{moving}$}
\psfrag{string1}{$\mbox{string}$}
\psfrag{galaxy}{$\mbox{source}$}
\psfrag{lensed}{$\mbox{lensed}$}
\psfrag{image}{$\mbox{image}$}
\psfrag{obs}{$\mbox{observer}$}
\includegraphics[width=3.2in]{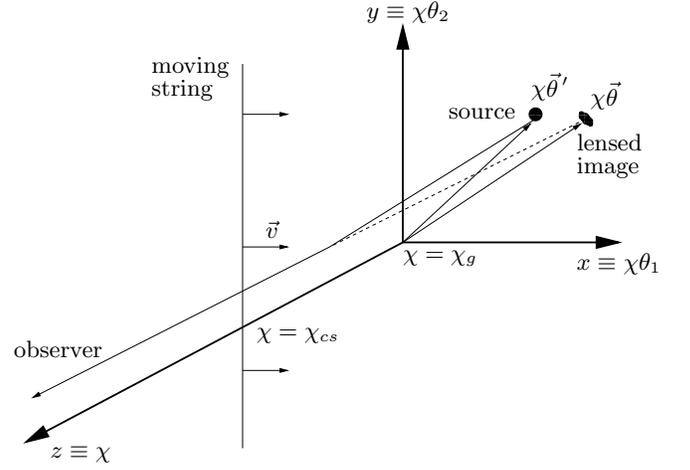}
\end{center}
\caption{The geometrical setup for the single string lensing
  calculation. The moving string is aligned with the $y$-axis,
  perpendicular to the line of sight. A light ray sourced at position $\chi\vec
  \theta^{\, \prime}$ in the source plane ($\chi=\chi_g$) is imaged onto $\chi\vec
  \theta$. The vector source due to the moving string can
  rotate an image in addition to the usual shearing for the general
  case where the string is not aligned with the $y$-axis.}
\label{fig:setup}
\end{figure}

{\it Deflection pattern around a single string.}
For simplicity the straight string is assumed to be 
aligned with the orthogonal
frame of reference as shown in Fig.~\ref{fig:setup},  moving with
velocity $v_{cs}$ perpendicular to the line of sight at a distance
$\chi_{cs}$. The source being lensed is placed behind the string at a
distance $\chi_g$ at position $ x=\chi \theta_1$, $y=\chi \theta_2$ in the
orthogonal plane. The vector string velocity (or momentum) 
$\vec v=(v_{cs},0,0)\delta(z-\chi_{cs})\delta(x)$ 
is composed of irreducible scalar and vector components $v^S$ and
${\vec \omega}$, with $\vec v = \vec\nabla v^S + \vec \omega$, where
$\vec \omega$ is divergenceless and $\vec \nabla \cdot \vec v =
\nabla^2v^S$. Thus in Fourier space the vorticity component takes the
form 
\begin{equation}\label{omega}
  \vec \omega = \vec v - \frac{\vec k}{k^2}\left( \vec{ k}\cdot \vec v\right)
=2\pi v_{cs} e^{-ik_z \chi_{cs}}\delta(k_y)
(\hat k_z^2,0,-\hat k_z\hat k_x)\,.
\end{equation}
Given the geometry of the setup only the $11$ component of the
distortion tensor is sourced by the $\omega$ and no rotation is
induced in the lensed image. The result is generalised to a string
aligned in a general direction by rotating the tensor once the
component $\psi_{11}$ has been obtained. A contribution to $\rho$ is
generated upon rotation to the general configuration where the string
is not aligned with the $y$-axis.

Substituting (~\ref{omega}) into (~\ref{psi}) and
neglecting the time--suppressed $\frac{\dot a }{a} \omega$ term, and
the subdominant $\Pi$ term \cite{thesis} gives
\begin{eqnarray}\label{psi11}
\psi_{11}&=&-\frac{4G v_{cs}}{\pi}\int^{\chi_g}_0 d\chi \chi\left(1-\frac{\chi}{\chi_g}
\right)\times \nonumber\\ 
&&\int^{\infty}_{-\infty}d^3k\,e^{i\left(\vec{k}\cdot\vec{x}-k_z \chi_{cs}\right)}\delta(k_y)\hat k_z \hat k_x\,.
\end{eqnarray}
Replacing $\hat k _i$ with its coordinate space equivalent
$\nabla_i$, (~\ref{psi11}) can be integrated  by parts to obtain
\begin{eqnarray}
\psi_{11}\!\!&=&\!\!-\frac{8G v_{cs}}{\xi^2_+}\left\{\zeta\xi_-\left[\tan^{-1}(\frac{1}{\t_1})+\tan^{-1}\left(\frac{\xi_+-\zeta}{\zeta\t_1} \right) \right]+\right.\nonumber \\ 
&&\left.\t_1\xi_++\frac{\zeta \t_1} {\xi_+}\log \left(\frac{\xi_+}{\zeta^2}-\frac{2}{\zeta}+1 \right) \right\}\,,
\end{eqnarray}
with boundary terms vanishing, $\zeta\equiv \chi_{cs}/\chi_g$ and $\xi_\pm
\equiv (1\pm \t_1^2)$. This solution maps the only non vanishing term
of the distortion matrix for a single string aligned with the $y$-axis
as a function of the string position relative to the source and the
transverse angular distance from the string. It is valid in the weak
lensing $\psi_{11}\ll 1$ and small angle $\theta_1 \ll 1$ regime. The
rotation $\rho$ arising from the general case where the string is
moving in a direction which is {\sl not} aligned with the $\theta_1$
axis is obtained by rotating the solution by an angle $\alpha$ to give $\rho = -\psi_{11}\sin
(\alpha)/2$.

\begin{figure}[t]
\begin{center}
\psfrag{xvar}{$\ \ \ \ \ \ \theta_1$}
\psfrag{ysvar}{$\ \ \zeta$}
\includegraphics[width=3.2in]{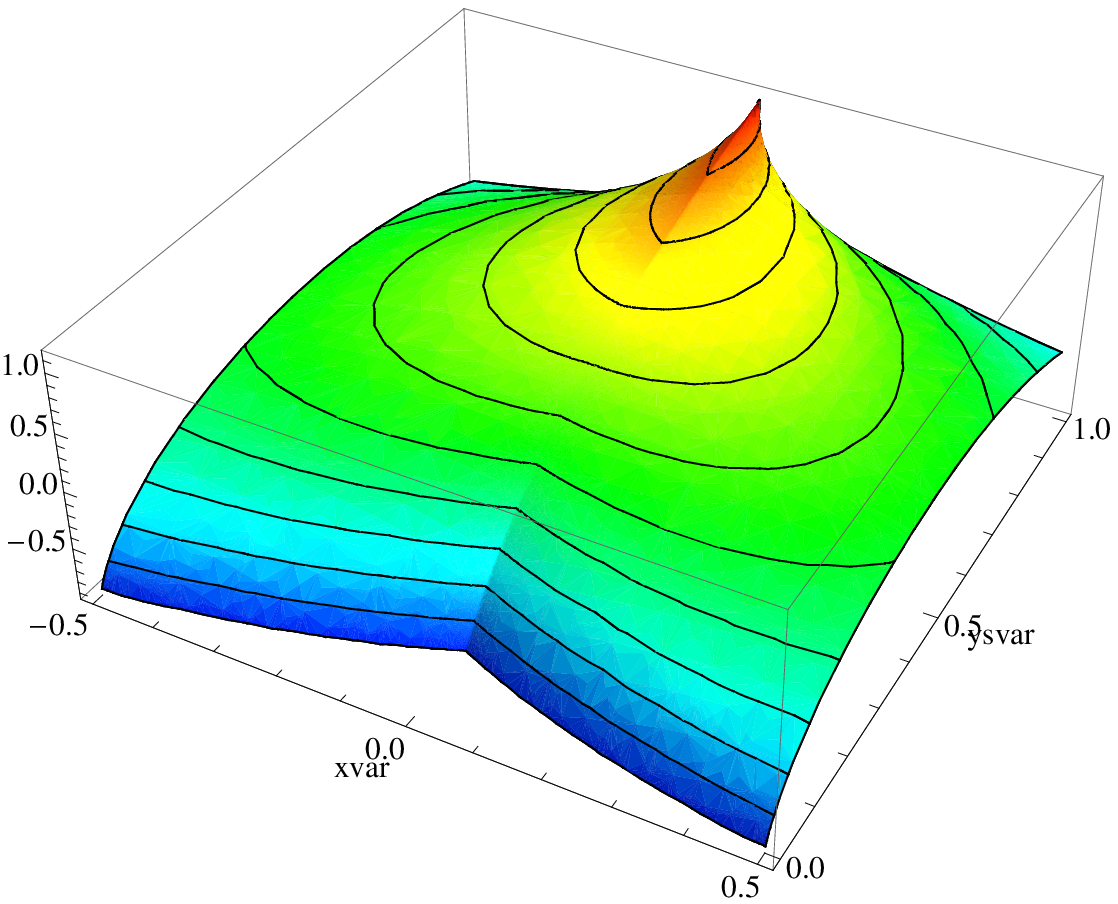}
\end{center}
\caption{The solution $\log |\rho|$, in arbitrary units, as a function
  of angular distance $\theta_1$ from the string and distance of the
  string with respect to the distance of the source object $\zeta\equiv
  \chi_{cs}/\chi_g$. The solution is only valid for $\theta_1 \ll
  1$. $\rho$ is anti-symmetric about $\theta_1=0$.}\label{fig:bcp}
\end{figure}

\begin{figure}[t]
\begin{center}
\includegraphics[width=2.6in,angle=270]{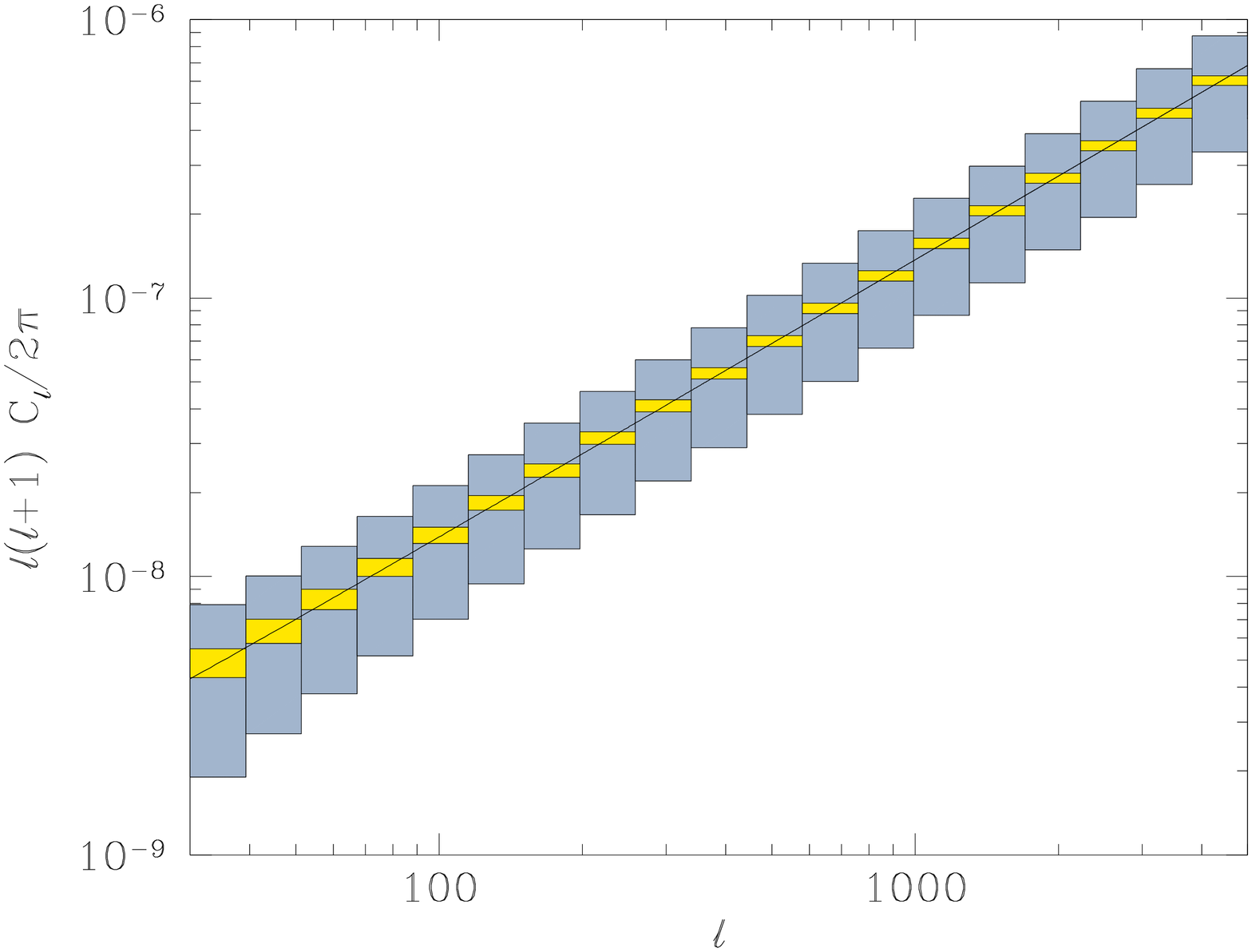}
\end{center}
\caption{The angular power spectrum of rotation for a network of
  strings with $G\mu = 1\times 10^{-7}$. The blue and yellow boxes
  show the forecasted error for two surveys with $f_{\rm sky}=0.1$ and
  $f_{\rm sky}=0.5$ respectively. 
  The
  errors include both sample and intrinsic elipticity noise
  contributions. The intrinsic elipticity term dominates at the
  relatively small scales being considered.}\label{fig:spectrum}
\end{figure}

The solution for $-0.5 \le \t_1 \le 0.5$ is shown in
Fig.~\ref{fig:bcp}. Whilst the rotation peaks in the limit $\zeta\sim 1$
with a steep drop-off in the transverse direction, it extends furthest
in $\theta_1$ when $\zeta\sim 1/2$ i.e. the case where the string is
placed midway between the source and observer. For this limit the
solution is approximated by
\begin{equation}
  \rho \approx 2\pi G v_{cs} \sin \alpha\left(1-3\t_1^2 \right)\,, 
\end{equation}
to second order in $\t_1$.

{\it Vector power spectrum.} In the presence of a network of strings
the signal must be calculated in terms of power spectra averaged over
the sky.  In this case the signal is assumed to be 
generated by a scaling network of cosmic strings with tension $\mu$ with the
limit $G\mu < 10^{-6}$ set by the allowed contribution to the scalar
angular power spectrum of the Cosmic Microwave Background
\cite{bevis1}. In the small angle limit the quantity of interest is
the 2d power spectrum of $\psi _{ij}$, $\langle  \psi _{ij}(\vec
l) \ \  \psi ^\star_{lm}(\vec {l^{'}}) \rangle=(2\pi)^2 \delta^2(\vec l
-\vec {l^{'}})P^{\psi}_{ijlm}(l)$, where $\vec l$ is the 2d fourier
transform reciprocal of $\vec \theta$. The 2d power spectrum for the
rotation is then 
\begin{eqnarray}\label{eq:prho}
P_{\rho}(l)\!\!\!&=&\!\!\!\int^{\c_{\infty}}_0\!\!\!\!\!\!d\c \frac{g^2(\c)}{\c^3}64\pi^2G^2 \times \nonumber \\
&&\left(\frac{4\dot{a}^2\c^2}{a^2l^2}P_{\omega}(l)+\frac{P_{\Pi}(l)}{4}+\frac{2\dot{a}\c}{al}P_{\Pi\omega}(l) \right)\,,
\end{eqnarray}
where the power spectra for the source terms $P_\omega(l)$,
$P_{\Pi}(l)$, and their cross--correlation $P_{\Pi\omega}(l)$ in the
small angle limit ($k_z \ll l/\chi$) have been introduced.

The source spectra for scaling networks of cosmic strings can be
written in terms of structure functions $P_X(k\chi,k\chi')$ which have
been measured from numerical simulations
\cite{Allen97,contaldi99,bevis2} and computed from semi-analytical
models \cite{Albrecht99,pogosian99}. The unequal time correlators for
the source terms are related to the structure functions through
scaling laws
\begin{equation}
  \langle \omega_i(\vec k,\e) {\omega}^\star_j(\vec {k^{'}},\e^{'})
  \rangle=(2\pi)^3\delta^3(\vec k - \vec
  {k^{'}})P_{ij}\frac{P_{\omega}(k\chi,k\chi^{'})}{\sqrt{\chi\chi^{'}}}
\,,
\end{equation}
with projector $P_{ij}=\delta_{ij}-\hat {k}_i \hat {k}_j$ and similar
relations for $\langle\Pi\Pi^\star\rangle$ and $\langle\Pi\omega^\star\rangle$ correlations. For
this case, in the small angle limit, only the diagonal of the structure
functions is relevant with $P_\omega(k\chi,k\chi^{'}) \rightarrow
P_\omega(l)$ where $l\approx k\chi$.

To determine whether a cosmic
string network could potentially be detected, (\ref{eq:prho}) can be
computed numerically. A simple power law description for the structure
functions can be adopted. This is justified by causality requirements
and the relative normalisations of the different correlations can be
taken from the numerical results of \cite{thesis,iran,bevis2}. An inverse
scaling with $l$ is assumed such that
\begin{equation}
  P_\omega(l) = \left( G\mu\right)^2 l^{-1}\,,
\end{equation}
with causality imposed 
relative normalisations $\omega:\Pi:\Pi\omega = 1:0.25:0.1$.
An overall amplitude $(G\mu)^2=10^{-14}$ is used throughout.

Fig.~\ref{fig:spectrum} shows the power spectrum of the rotation
$C_\ell^\rho$ where in the small angle limit $\ell\approx l$ and
$C_\ell^\rho \sim (2\pi)^2 P_\rho(l)$. The integral in
(\ref{eq:prho}) is computed assuming a background galaxy distribution as a
function of redshift $z$ as $w(z)\sim z^2 \exp(-z/z_0)$ with $z_0=0.4$
and taking the maximum redshift to be $z=6$. Cosmological
parameters $\Omega_m=0.3$, $\Omega_{\Lambda}=0.7$ and
$h=0.72$ are used. Expected errors for two surveys covering 10\% and 50\% of
the sky are also shown. The errors include contributions from both
sample and intrinsic elipticity noise variance although the latter
dominates the errors at these scales. For both surveys a
background galaxy density of 100 galaxies per square arcminute and an
average intrinsic elipticity of 0.3 was assumed.

As shown a distinct weak lensing signal generated by cosmic strings is
predicted: rotation with a specific power spectrum. Its intensity is
below current sensitivity but provides an ideal target for projected
observations. If cosmic string networks exist with $G\mu \sim 10^{-7}$
then the effect should be detectable with the next generation of
surveys \cite{lsst,panstarrs}.  Should it not be observed then the
constraints on a string network will become considerably tighter.  The
only caveat is that at this level we can no longer assume that no
curl--like modes are generated by lensing.  Separating the
corresponding systematics out in these surveys will therefore be more
challenging. Yet, the distinct spectral signature of the string signal
is likely to provide a simple solution to this problem.

{\sl Acknowledgements} This work was supported by an STFC studentship.

\bibliography{something}
\end{document}